\documentclass[11pt,a4paper]{article}
\usepackage{jheppub}
\usepackage{graphicx}
\usepackage{amssymb}
\usepackage{amsmath}
\usepackage{topcapt}
\usepackage{epstopdf}
\title{Near commuting multi-matrix models}

\author[a]{ Denjoe O'Connor}
\author{and}
\author[b]{Veselin G. Filev}

\affiliation[a,b]{School of Theoretical Physics, Dublin Institute for Advanced Studies\\
10 Burlington Road, Dublin 4, Ireland.}
\emailAdd{denjoe@stp.dias.ie}
\emailAdd{vfilev@stp.dias.ie}

\abstract{We investigate the radial extent of the eigenvalue distribution
for Yang-Mills type matrix models.  We show that, a three matrix 
Gaussian model with complex Myers coupling, to leading order in
strong coupling is described by commuting matrices whose joint
eigenvalue distribution is uniform and confined to a ball of radius
$R={\left(\frac{3\pi}{2g}\right)}^{1/3}$.  We then study, perturbatively, 
a $3$-component model with a pure commutator action and find a 
radial extent broadly consistent with numerical simulations.}


\subheader{}

\begin{document}
\maketitle

\section{Introduction}

Multi-matrix models arise in a wide variety of settings 
from matrix string theory \cite{Dijkgraaf:1997vv}, the IKKT model
\cite{Ishibashi:1996xs} (and its lower dimensional variants
\cite{Connes:1997cr}), the BFFS and BMN models
\cite{Banks:1996vh,Berenstein:2003gb} to the low energy dynamics of
$D$-branes \cite{Kazakov:1998ji} and simple models of emergent
geometry \cite{DelgadilloBlando:2007vx} and emergent gravity 
\cite{Steinacker:2012ra,Blaschke:2010ye}. 
More recently the BMN and BFFS models have
received attention in the context of fast scrambling in
\cite{Asplund:2012tg}.

In the context of numerical simulations it has been observed that the
$3$-component Yang Mills matrix model (with pure commutator action) 
\cite{DelgadilloBlando:2012xg} has an eigenvalue distribution that is 
parabolic with radial extent $R=2.0$. It is tempting to suspect that
this result should be derivable from first principals and this paper is 
an attempt in this direction. In part the motivation for the current paper was 
to attempt a theoretical estimate of the extent of the eigenvalue distribution
in this simple model. Though we did not succeed in getting
an exact result we succeeded in getting approximate 
results which are in broad agreement with the simulations. 

The principal results of this paper are:
\begin{itemize}
\item A derivation of the uniform ball distribution in a previously studied 
3-matrix model, some of whose observables have been solved exactly.
\item A demonstration that one can obtain the exact extent of the
  eigenvalue distribution in this 3-matrix model from an effective
  potential for this radial extent.
\item The derivation of an effective potential to 2-loops for the pure 
Yang-Mills matrix model.
\item We derive the measure for a rotationally invariant gauge fixing which is 
natural if the matrices are approximately commuting.  
\item Estimates for the radial extent in the 3-matrix Yang-Mills model,
our estimate is that the extent of the eigenvalue distribution is 
$R=1.6\pm 0.5$. 
\end{itemize}

The structure of the paper is as follows:

Section 2 is dedicated to the mass regulated two matrix model. This
model was introduced in \cite{Hoppe:PhDThesis1982} and arose again in
the description of the low energy dynamics of $N$ $D$-branes which are
close together in $4$-dimensional space-time \cite{Kazakov:1998ji}. The
strong coupling properties of the model were investigated in
ref.~\cite{Kazakov:1998ji} and in \cite{Berenstein:2008eg} it was
shown that at strong coupling the model is in a commuting phase with
the joint eigenvalue distribution described by a hemisphere.

In the first part of Section 2 we review the results of
ref.~\cite{Berenstein:2008eg} using a sightly different approach. In
particular we split the matrices into their diagonal and off-diagonal
elements and consider a general axial gauge which is equivalent to
diagonalizing one of the matrices. We integrate out the perpendicular
modes and obtain an effective action for the diagonal components of
the matrices. Next we consider a coarse grained approximation and show
that the longitudinal part of the diagonal modes has a parabolic
eigenvalue distribution~\cite{Berenstein:2008eg}.

In the second part of Section 2 we consider an $SO(3)$ invariant
non-Hermitian three matrix model constructed from a Hermitian mass
term and an anti-Hermitian Myers term \cite{Myers:1999ps}.
Integrating out one of the matrices one recovers the $SO(2)$ invariant
two matrix model considered in the first part of Section 2. However,
we study the model directly in the three matrix model realization by
splitting the matrices into their diagonal and off-diagonal components
and consider a general axial gauge for the off-diagonal modes. Next we
integrate out the perpendicular modes and obtain a one-loop effective
action (exact in the axial gauge) for the longitudinal part of the
diagonal components. We average over all possible orientations of the
constant unit vector specifying the axial gauge and obtain a three
dimensional $SO(3)$ invariant effective action for the diagonal
modes. Next we study the strongly coupled regime of the model and show
that to leading order the ground state is described by a uniform joint
eigenvalue distribution inside a solid ball. The hemisphere and
parabolic distributions conjectured in ref.~\cite{Berenstein:2008eg}
can then be obtained by integrating the uniform distribution over one
or two of the three coordinates respectively. Further, by conjecturing
a uniform joint eigenvalue distribution, we obtain an effective
potential for the radius of the distribution, which in the strong
coupling limit recovers the exact radial extent of the
distribution. It is this uniform joint eigenvalue distribution of the
model which motivates the studies presented in Section~3 of the
paper. We establish in Appendix \ref{AppendixA2} that there is a
unique lift of a one dimensional distribution to a rotationally
invariant three dimensional distribution, and therefore the unique
lift of the parabola is the uniform distribution. This can further be
reduced to a unique rotationally invariant two dimensional
distribution. Also, we develop the first order corrections to the
leading large $g$ behaviour in Appendix \ref{AppendixA2} and verify
that they are sufficient to capture the next term in a large $g$
expansion of an observable whose exact result is know.

Section 3 of our work explores the properties of the $SO(p)$ invariant
Hermitian p-matrix model corresponding to a commutator squared term,
i.e. a pure Yang-Mills type matrix model.  This family of models have
been studied both numerically and in a large $p$ expansion in
\cite{Hotta:1998en}. Special attention was given to the $p=3$ case in
ref.~\cite{Azuma:2004ie,O'Connor:2006wv,DelgadilloBlando:2008vi,Azuma:2004zq,Azuma:2005bj}, including
the effect of finite mass and Myers terms.  In particular
it has been shown that, as the strength of the Myers coupling is
varied, the ground state of the model undergoes an exotic phase
transition from a fuzzy two sphere to a matrix phase \cite{DelgadilloBlando:2007vx}. It is this matrix phase that we
explore in Section 3.

Section 3 contains four subsections.  In first part of the Section
3 we outline a change of coordinates involving splitting the matrices
to diagonal and off-diagonal components. The change of coordinates
that we consider is also equivalent to an $SO(p)$ invariant gauge
fixing condition for the off-diagonal elements. We calculate the
corresponding Jacobian to second order in the off-diagonal
modes. This takes into account contribution of the corresponding ghost
modes to two loops. At tree level the ghost determinant is a generalized $SO(p)$
invariant Vandermonde type determinant.

The second part of Section 3 derives the effective action for the
diagonal modes.  By setting up a systematic perturbative expansion we
obtain a two loop effective action for the diagonal elements of the
model and under the assumption of an $SO(p)$ invariant distribution
for these modes, within a ball of radius $R$, we obtain an effective
potential for the radial extent of their distribution.  We show that
the radial extent of these modes stabilizes in this approximation. For
the three matrix model assuming a uniform distribution inside a solid
ball, which by appendix \ref{AppendixA2} is the unique rotational
invariant lift of a parabolic one dimensional distribution, we obtain
an analytic estimate of the radius of the distribution.

The third part of Section 3 studies the gauge dependence of our
perturbative expansion. We consider a standard Faddeev-Popov gauge
fixing procedure and compare the results obtained in Feynman and
Landau gauges. We argue that for a generic gauge the joint
distribution of the diagonal modes will differ from the genuine joint
distribution and that the Landau gauge is the most suitable
for studying the model.

Section 4 of our paper is our concluding section. We summarize our
results and discuss possible directions for future studies.

\section{Strongly coupled mass regulated two matrix model}
Let us consider the Hermitian two matrix model first considered by Hoppe \cite{Hoppe:PhDThesis1982}:
\begin{equation}
{\cal Z}=\int{\cal D}X{\cal D}Ye^{-N{\rm{tr}}(X^2+Y^2-g^2{[X,Y]}^2)}\ .\label{partfunct2}
\end{equation}

The matrix model (\ref{partfunct2}) arose again in the work of Kazakov
et al \cite{Kazakov:1998ji} where the partition function for large $g$
and a fresh derivation of Hoppe's solution for the observable
$<\frac{\rm{tr}}{N}(X^2)>$ are given. The solution for these
quantities are given in a rather complex form, which makes a more
complete analysis of the theory (valid for large $g\gg 1$)
difficult. A more detailed understanding of the model was initiated in
ref.~\cite{Berenstein:2008eg} where it was suggested that the joint
eigenvalue distribution of the commuting saddle is given by a
hemisphere distribution. The analysis of
ref.~\cite{Berenstein:2008eg} shows that at large coupling the
eigenvalues of a single matrix are described by a parabolic eigenvalue
distribution.  The authors then argue that this suggests a hemisphere
eigenvalue distribution for the commuting saddle of the two matrix
model.


\subsection{Hemisphere distribution}
Let us review the results of ref.~\cite{Berenstein:2008eg}, by
following a slightly different approach which we find instructive. In particular we split the matrices to diagonal and
off-diagonal components and integrate out the off-diagonal modes by
imposing an axial gauge condition. To begin with let us define:
\begin{equation}
X_{ij}=x^1_i\delta_{ij}+a^1_{ij};~~~Y_{ij}=x^2_i\delta_{ij}+a^2_{ij};~~~\vec x_i=(x^1_i,x^2_i);~~~\vec a_{ij}=(a^1_{ij},a^2_{ij})\ .
\end{equation}
The action in (\ref{partfunct2}) can be written as:
\begin{equation}
S[X,Y]=\sum_{i}{\vec x_i}^2+\sum_{i\neq j}|{\vec a_{ij}}|^2+g^2\sum_{i\neq j}|(x^\mu_i-x^\mu_j)a_{ij}^{\nu}\epsilon_{\mu\nu}+(a^{\mu}a^{\nu})_{ij}\epsilon_{\mu\nu}|^2;~~~\mu,\nu=1,2\ .
\end{equation}
Let us also consider a constant unit vector $\vec n=(n^1,n^2)$ and define:
\begin{equation}
\vec a^{||}=\vec n(\vec n.\vec a);~~~\vec a^{\perp}=(\hat 1-\vec n\vec n).\vec a\ .
\end{equation}
Now we can use the $SU(N)$ symmetry of the matrix model to fix the gauge:
\begin{equation}
\vec n.\vec a_{ij}=0\ .\label{gauge}
\end{equation}
Note that in this gauge the action is quadratic in the off-diagonal modes and hence the first loop effective action is exact! Indeed, without loss of generality we can consider $n^1\neq0$ and express:
\begin{equation}
a^1=-\frac{n^2}{n^1}a^2\ .
\end{equation}
It is then easy to check that:
\begin{equation}
a^{\mu}a^{\nu}\epsilon_{\mu\nu}=[a^1,a^2]=-\frac{n^2}{n^1}[a^2,a^2]=0\ .
\end{equation}
This simplification is not a surprise since the gauge condition (\ref{gauge}) with a  constant $\vec n$ is equivalent to diagonalizing one of the matrices. The simplified action can be written as:
\begin{equation}
S[X^{\mu}]=\sum_{i}\left[(\vec n.\vec x_i)^2+{{\vec x}_i^{\perp2}}\right]+\sum_{i\neq j}\left[1+g^2(\vec n.(\vec x_i-\vec x_j))^2\right]|{\vec a^{\perp}_{ij}}|^2+S_{\rm{FP}}\ .\label{simple}
\end{equation}
Here $S_{FP}$ is the contribution from the Faddeev-Popov determinant resulting from the gauge choice (\ref{gauge}). We have also split the diagonal modes into longitudinal $\vec n.\vec x$ and perpendicular ${\vec x^{\perp}}_i$ components. It is easy to verify that:
\begin{equation}
S_{FP}=-\frac{1}{2}\sum_{i\neq j}\log(\vec n.(\vec x_i-\vec x_j))^2\ .
\end{equation}
Note that if we choose a basis in which $\vec n=(1,0)$ or $\vec n=(0,1)$ the Faddeev-Popov determinant is just the standard Vandermonde determinant. It is now straightforward to integrate out the perpendicular elements of the matrices. The resulting effective action (we divide by $N^2$) for the longitudinal diagonal modes is:
\begin{equation}
S_{\rm{eff}}[\vec x]=\frac{1}{N}\sum_{i}({\vec n.\vec x_i})^2-\frac{1}{2N^2}\sum_{i\neq j}\log\left[\frac{(\vec n.(\vec x_i-\vec x_j))^2}{1+g^2(\vec n.(\vec x_i-\vec x_j))^2}\right]\ .\label{eff-action}
\end{equation}
A few comments are in order: Note that the first loop effective action
(\ref{eff-action}) is exact in this gauge. It is also valid for all
$g^2$. However, we know that the matrices commute only in the large
$g^2$ limit, so the meaning of the individual components of
$\vec{x}_i$ needs clarification. Our strategy will be to lift the
eigenvalue distribution of the matrix $\vec{n}.\vec{X}$ specified by
(\ref{eff-action}) to a rotationally invariant 2-dimensional
distribution which in the large $g$ limit becomes the joint eigenvalue
distribution for the commuting eigenvalues of $X$ and $Y$.  See
\ref{AppendixA2} for a more detailed discussion.

In this spirit we consider a coarse grained
description and extremize the following functional:
\begin{eqnarray}
S_{\rm{eff}}[\rho(\vec x)]&=&\int d^2x\rho(\vec x)(\vec n.\vec x)^2-\frac{1}{2}\int\int d^2x d^2x'\rho(\vec x)\rho(\vec x')\log\left[\frac{(\vec n.(\vec x-\vec x'))^2}{1+g^2(\vec n.(\vec x-\vec x'))^2}\right]+\quad\quad\quad \\
&+&\mu\left(\int d^2x\rho(\vec x)-1\right)\nonumber
\end{eqnarray}
Upon variation with respect to $\rho$ we obtain\footnote{If we average over $n$ here we obtain
$$\mu+\frac{\vec{x}^2}{2}=2\int d^2x'\rho(\vec{x'})\ln(\frac{\vert\vec{x}-\vec{x'}\vert}
{1+\sqrt{1+g^2\vert\vec{x}-\vec{x'}\vert^2}}) .$$} the equation:
\begin{equation}
\mu+(\vec n.\vec x)^2=\int d^2x'\rho(\vec x')\log\left[\frac{(\vec n.(\vec x-\vec x'))^2}{1+g^2(\vec n.(\vec x-\vec x'))^2}\right]\ .\label{EQN}
\end{equation}
Now we can apply the following differential operator $\vec n.\nabla_{\vec x}$ to both hand sides of equation (\ref{EQN}). The result is:
\begin{equation}
\vec n.\vec x=\int \frac{d^2x'\rho(\vec x')}{(\vec n.(\vec x-\vec x'))(1+g^2(\vec n.(\vec x-\vec x'))^2)}
\end{equation}
This is the equation that we can use to deduce the form of the joint eigenvalue distribution $\rho(\vec x)$. Note that this is almost equation (5) from ref. \cite{Berenstein:2008eg}. To make the analogy complete we define $u=\vec n.\vec x$ and choose a coordinate frame in the integral along $\vec x'$ in which we have $\vec n.\vec x'=x'^1$.
\begin{equation}
u=\int \frac{dx'^1\rho_1( x'^1)}{(u-x'^1)(1+g^2(u-x'^1)^2)}\label{Eqn2}
\end{equation}
Note that we have defined:
\begin{equation}
\rho_1(x^1)=\int dx^2\rho(x^1,x^2)\label{Eqn11}
\end{equation}
We next observe that for large $g$:
\begin{equation}
\frac{1}{2}\log\left[\frac{g^2x^2}{1+g^2x^2}\right]=-\frac{\pi}{g}\,\delta(x)+O(1/g^2)\quad\hbox {and}\quad \frac{1}{ x (1+g^2x^2)}=-\frac{\pi}{g}\delta'(x)+O(1/g^2)\ ,\label{Delta-conv}
\end{equation}
Substituting in equation  (\ref{Eqn2}) we obtain:
\begin{equation}
u=-\frac{\pi}{g}\int{dx'^1\rho_1( x'^1)}\delta'(u-x'^1)+O(1/g^2)=-\frac{\pi}{g}\rho_1'(u)+O(1/g^2)\ .\label{eqn4}
\end{equation}
Next we solve equation (\ref{eqn4}) to leading order in $g^{-1}$ and normalize $\int du\rho_1(u)=1$. The resulting distribution is given by \cite{Berenstein:2008eg}:
\begin{equation}
\rho_1(x^1)=\frac{3}{4R^3}(R^2-{x^1}^2)\ ,\label{parabolic}
\end{equation}
with 
\begin{equation}
R={\left(\frac{3\pi}{2g}\right)}^{1/3}\ .\label{radL}
\end{equation}
Now the definition of $\rho_1$ from equation (\ref{Eqn11}) and the $SO(2)$ symmetry of the distribution together with Appendix \ref{AppendixA2} gives  the unique solution for $\rho_2(x^1,x^2)$:
\begin{equation}
\rho_2(x^1,x^2)=\frac{3\sqrt{R^2-\vec x^2}}{2\pi R^3}\ , \label{hemi-sphere}
\end{equation}
as originally suggested in \cite{Berenstein:2008eg}.
Equation (\ref{hemi-sphere}) is the desired hemisphere distribution reported in ref.~\cite{Berenstein:2008eg}. Note that the derivation of the two-dimensional distribution (\ref{hemi-sphere}) is somewhat indirect. Indeed what we derived from the effective action (\ref{eff-action}) was the one-dimensional parabolic distribution (\ref{parabolic}). In the next subsection we will consider a three matrix model equivalent to the two matrix model (\ref{partfunct2}) and prove directly that at strong coupling the corresponding three dimensional eigenvalue distribution is an uniform distribution inside a solid ball. The two dimensional hemisphere eigenvalue distribution can be obtained by integrating out one of the eigenvalues.


\subsection{Three matrix model realization and uniform distribution}

Let us now consider the model (also originally introduced parenthetically 
by Hoppe \cite{Hoppe:PhDThesis1982} page 73 and further discussed in 
\cite{Kazakov:1998ji}):
\begin{equation}
{\cal Z}=\int{\cal D}X{\cal D}Y{\cal D}Ze^{-N\rm{tr}(X^2+Y^2+Z^2-i\alpha [X,Y]Z)}\ .\label{partfunct3}
\end{equation}
It is easy to verify that if one integrates out the $Z$ matrix and defines $g^2={(i\alpha)^2}/{4}$ one recovers the two matrix model (\ref{partfunct2}). 
This suggests that the model (\ref{partfunct3}) should be as solvable 
as the two matrix model. Note also that there is a global $SO(3)$ symmetry rotating the $X,Y$ and $Z$ matrices. We find it instructive to analyze the model in the spirit described in section 2.1. To begin with let us define:
\begin{equation}
X_{ij}=x^1_i\delta_{ij}+a^1_{ij};~~Y_{ij}=x^2_i\delta_{ij}+a^2_{ij};~~Z_{ij}=x^3_i\delta_{ij}+a^3_{ij};~~\vec x_i=(x^1_i,x^2_i,x^3);~~\vec a_{ij}=(a^1_{ij},a^2_{ij},a^3_{ij})\ .
\end{equation}
Next we consider a constant unit vector $\vec n=(n^1,n^2,n^3)$, define:
\begin{equation}
\vec x^{\perp}=(\hat 1-\vec n\vec n).\vec x;~~~\vec a^{||}=\vec n(\vec n.\vec a);~~~\vec a^{\perp}=(\hat 1-\vec n\vec n).\vec a\ .
\end{equation}
and impose the axial gauge $\vec n.\vec a=0$. The action in (\ref{partfunct3}) can then be written as:
\begin{equation}
S[\vec x,\vec a]=\sum_{i}\left[(\vec n.\vec x_i)^2+{{\vec x}_i^{\perp2}}\right]+\sum_{i\neq j}a^{\mu\perp}_{ij}\left[\delta^{\mu\nu}-i\frac{\alpha}{2}\epsilon_{\mu\nu\rho}n^{\rho}(\vec n.\vec\Delta_{ij})\right]a^{\nu\perp}_{ji}\ ,\label{act3}
\end{equation}
where we have defined $\vec\Delta_{ij}=\vec x_i-\vec x_j$. Note that there is no term cubic in $\vec a$ in the action (\ref{act3}), because of the axial gauge. This means that the one loop effective action is exact. Now we proceed as in section 2.1 and integrate the perpendicular matrix elements. One can show that the resulting effective action is given by:
\begin{equation}
S_{\rm{eff}}[(\vec n.\vec x)]=\frac{1}{N}\sum_{i}({\vec n.\vec x_i})^2-\frac{1}{2N^2}\sum_{i\neq j}^N\log\left[\frac{g^2 (\vec n.(\vec x_i-\vec x_j))^2}{1+g^2(\vec n.(\vec x_i-\vec x_j))^2}\right]+\frac{(N-1)}{2N}\log g^2\ ,\label{eff-action3}
\end{equation}
where $g^2=(i\alpha)^2/4$. Next we consider a coarse grained approximation and vary the corresponding distribution function $\rho$ to obtain the equation:
\begin{equation}
\mu+(\vec n.\vec x)^2=\int d^3x'\rho(\vec x')\log\left[\frac{g^2(\vec n.(\vec x-\vec x'))^2}{1+g^2(\vec n.(\vec x-\vec x'))^2}\right]\ .\label{EQN3}
\end{equation}
Note that equation (\ref{EQN3}) is valid for any choice of $\vec n$. Next we average over all possible directions that $\vec n$ can take with a uniform weight. Or equivalently average over the unit two-sphere. It is easy to show that:
\begin{equation}
\frac{1}{4\pi}\int d\Omega_2(\vec n.\vec x)^2=\frac{1}{3}\vec{x}^2\ . \label{EQNAVX} 
\end{equation}
The right-hand side of equation (\ref{EQN3}) requires a bit more careful analysis. One can show that:
\begin{eqnarray}
&&J(g|\vec{x}-\vec{x'}|):=\frac{1}{4\pi}\int d\Omega_2\log\left[\frac{g^2(\vec n.(\vec x-\vec x'))^2}{1+g^2(\vec n.(\vec x-\vec x'))^2}\right]=\frac{1}{2g|\vec x-\vec x'|}\int\limits_{-g|\vec x-\vec x'|}^{g|\vec x-\vec x'|}d\eta\log\left(\frac{\eta^2}{1+\eta^2}\right)\nonumber\\
&&=\frac{-2\arctan(g|\vec x-\vec x'|)}{g|\vec x-\vec x'|}+\log\left[\frac{g^2(\vec x-\vec x')^2}{1+g^2(\vec x-\vec x')^2}\right]=-\frac{\pi}{g}\frac{1}{|\vec x-\vec x'|}+O\left(\frac{1}{g^2}\right)\ ,\label{KernelExpansion}
\end{eqnarray}
where the term $O(g^{-2})$ should be thought of as the difference of the lefthand side of equation (\ref{KernelExpansion}) and the $g^{-1}$ term on the righthand side of the equation.
To leading order in $1/g$ equation (\ref{EQN3}) becomes:
\begin{equation}
\mu+\frac{1}{3}\vec x^2=-\frac{\pi}{g}\int d^3x'\rho(\vec x')\frac{1}{|\vec x-\vec x'|}\label{EQAV}\ .
\end{equation}
Next we apply the Laplacian $\Delta_{x}$ on both hand-sides of equation (\ref{EQAV}) to obtain:
\begin{equation}
2=\frac{4\pi^2}{g}\int d^3x'\rho(\vec x')\delta(\vec x-\vec x')=\frac{4\pi^2}{g}\rho(\vec x)\ .
\end{equation}
Therefore we conclude that:
\begin{equation}
\rho(\vec x)=\frac{g}{2\pi^2}={\rm const}\label{rho}
\end{equation}
and hence at strong coupling we get a uniform eigenvalue distribution 
inside a ball. 
In order to estimate the radius of the eigenvalue distribution $R$ we use the normalization of $\rho(\vec x)$:
\begin{equation}
\int d^3x\rho(\vec x)=\frac{4\pi}{3}R^3\rho=1\ .\label{normalization}
\end{equation}
From equations (\ref{rho}) and (\ref{normalization}) we obtain:
\begin{equation}
R={\left(\frac{3\pi}{2g}\right)}^{1/3}\ , \label{radius}
\end{equation}
which is exactly the radius obtained in ref.~\cite{Berenstein:2008eg} reported in equation (\ref{radL}). It is also straightforward to obtain the hemisphere distribution (\ref{hemi-sphere}). Indeed:
\begin{equation}
\rho_2(x^1,x^2)=\int\limits_{-\sqrt{R^2-{x^1}^2-{x^2}^2}}^{\sqrt{R^2-{x^1}^2-{x^2}^2}}dx^3\rho(x^1,x^2,x^3)=\frac{3\sqrt{R^2-{x^1}^2-{x^2}^2}}{2\pi R^3}\ .
\end{equation}

Some additional comments are worthwhile here. First one can push the 
analysis further by observing that if we don't assume large coupling 
instead of (\ref{EQAV}) we obtain 
\begin{equation}
\mu+\frac{1}{3}\vec x^2=\int d^3x'\rho(\vec x')J(g|\vec{x}-\vec{x'}|).
\end{equation}
which upon acting with the Laplacian and noting 
\begin{equation}
\nabla^2J(x)=\frac{2}{(1+g^2|\vec{x}-\vec{x'}|^2)|\vec{x}-\vec{x'}|^2}
\end{equation}
we obtain the integral equation
\begin{equation}
1=\int d^3x'\frac{\rho(\vec x')}{(1+g^2|\vec{x}-\vec{x'}|^2)|\vec{x}-\vec{x'}|^2}\label{IE1}
\end{equation}
whose large $g$ behaviour is
\begin{equation}
\frac{1}{(1+g^2|\vec{x}-\vec{x'}|^2)|\vec{x}-\vec{x'}|^2}=\frac{\pi}{2g}\frac{\delta(\vert\vec{x}-\vec{x'}\vert)}{\vert\vec{x}-\vec{x'}\vert^2}+O(1/g^2)=
\frac{2\pi^2}{g}\delta(\vec{x}-\vec{x'})+O(1/g^2)\label{IE2}
\end{equation}
So to leading order we obtain $\rho(x)=\frac{g}{2\pi^2}\theta(R-r)$.

Given that the eigenvalue distribution is concentrated in the interior of 
a ball of radius $R$ we can further deduce that this radius is determined by
\begin{equation}
g^2=\frac{2}{3\pi}{(g R)}^3+\cdots \quad \hbox{and that} \quad \nu=\frac{g^2}{3}\int\rho(\vec{x})\vec{x}^2 =\frac{{(gR)}^2}{5}+\dots \label{nu2.2}
\end{equation}
from which we find to leading order in large $g$ the observable 
\begin{equation}
\nu=\frac{{(12\pi)}^{2/3}}{20}g^{4/3}+\cdots
\end{equation}
in agreement with \cite{Kazakov:1998ji}.

So far our considerations concerned only the leading order behaviour
of the distribution at large $g$. However, to verify the stability of
our result we have to explore the next order corrections to the
uniform distribution (\ref{rho}). It turns out that one can obtain an
integral equation of the second kind for the correction to the uniform
distribution $\Delta\rho$ in terms of a kernel 
$$\Delta K(\vec{x}-\vec{x'})=\frac{g}{4\pi^2}J(\vec{x}-\vec{x'})-\delta(\vec{x}-\vec{x'}).$$ This
equation can be developed recursively in a series providing a
perturbative expansion of the corrections to the uniform distribution
at large g. We refer the reader to  Appendix \ref{AppendixA1} for
a detailed derivation of the recursive series and calculation of the
next to leading order correction to the distribution.  In this
appendix we also check our result by calculating the next to leading
order term in the large $g$ expansion of the observable $\nu$ defined
in (\ref{nu2.2}). Again our results agree with the results of
\cite{Kazakov:1998ji}, see equation (\ref{A15}).

Further: Let us provide an alternative derivation of the radius of the distribution (\ref{radius}). If we average over $\vec n$ the effective action (\ref{eff-action3}), assume a constant eigenvalue distribution within a sphere of radius $R$ i.e. 
$\rho(r)=\frac{3\theta(R-r)}{4\pi R^3}$ and use 
\begin{equation}\int d^3xd^3x'\frac{\rho(x)\rho(x')}{\vert\vec{x}-\vec{x}'\vert}=\frac{6}{5 R}
\end{equation}
to leading order in $1/g$ we obtain:
\begin{equation}
S_{eff}=V_{eff}(R)=\frac{R^2}{5}+\frac{3\pi}{5gR}\ .
\end{equation}
Varying with respect to $R$ we find that $R^3=\frac{3\pi}{2g}$ in agreement with
the exact expression (\ref{radL}). In the next section we attempt to use such 
an effective potential (derived perturbatively) to estimate the extent of the
eigenvalue distribution in a three matrix model with pure commutator action.


\section{The $p\geq 3$ matrix model}
In this section we consider the p-matrix model:
\begin{equation}
S[X]=N Tr\left(-\frac{1}{4}[X^{\mu},X^{\nu}]^2\right);~~~\mu,\nu=1\dots p;\label{pmodel}
\end{equation}
where $X_a$ are Hermitian $N\times N$ matrices. The partition function is given by:
\begin{equation}
{\cal Z}=\int dX_a e^{-S[X]}=\int{\cal D}Xe^{N\frac{1}{4}Tr[X^{\mu},X^{\nu}]^2} .
\label{partfunct}
\end{equation}
Note that in addition to the $SU(N)$ gauge
invariance the model (\ref{partfunct}) has a global $SO(p)$ symmetry
transforming the matrices $X^{\mu}$.  We are interested in the eigenvalue distribution of one of the matrices and in particular in the extent $R$ of the 
eigenvalue distribution in the large $N$ limit. 
We find it convenient to split the degrees of freedom to diagonal and
off-diagonal contributions:
\begin{equation}
X^{\mu}_{ij}=x^{\mu}_i\delta_{ij}+a^{\mu}_{ij}.
\end{equation}
In terms of the new variables the action in (\ref{partfunct}) can be written as:
\begin{eqnarray}
S[X]=N\frac{1}{2}\sum_{i\neq j} |\vec\Delta_{ij}|^2{a}^{\mu}_{ij}\Pi^{\mu\nu}_{i j}a^{\nu}_{ji}-N\sum_{i\neq j}\Delta^{\mu}_{ij} a^{\nu}_{ij}[a^{\mu},a^{\nu}]_{ji}-N\frac{1}{4}tr[a^{\mu},a^{\nu}]^2 \ ,\label{actionsplit}
\end{eqnarray}
where we have defined:
\begin{equation}
\Delta^{\mu}_{ij}=x^{\mu}_i-x^{\mu}_j;~~~n^{\mu}_{ij}=\Delta^{\mu}_{ij}/|\vec\Delta_{ij}|;~~~ \Pi^{\mu\nu}_{ij}=\delta^{\mu\nu}-n^{\mu}_{ij}n^{\nu}_{ij};\ .\label{projector}
\end{equation} 
A standard way to proceed would be to integrate out the off-diagonal
degrees of freedom $a^{\mu}_{ij}$ and obtain an effective action for
the diagonal components $x^{\mu}_i$. Note that the quadratic term in
$a^{\mu}$ in equation (\ref{actionsplit}) is proportional to a
projector and hence cannot be directly inverted. A gauge fixing is
required. We find it natural to work in a gauge in which the
longitudinal modes are removed, more precisely we impose the gauge
fixing condition $\vec n_{ij}.\vec a_{ij}=0$. In the next subsection
we briefly present the change of variables necessary to implement our
gauge fixing condition. We refer the reader to Appendix B for a more
detailed calculation of the corresponding Jacobian.


\subsection{Gauge fixing}
Our goal is to perform a change of coordinates which is manifestly $SO(p)$ invariant and convenient in calculating quantum corrections to the effective potential governing the ground state of the theory.

Let us consider a set of $p-1$ Hermitian matrices ${a^m_{ij}}^{\perp}$ ($m=1\dots p-1$; $i,j=1\dots N$) with vanishing diagonal components (${a_{ii}^m}^{\perp}=0$). A slightly more general way to parametrize this set of matrices is to consider a set of $p$ linearly dependent matrices ${a^{\mu}_{ij}}^{\perp}$ ($\mu=1\dots p$) satisfying:
\begin{equation}
\sum\limits_{\mu=1}^pn^{\mu}.{a^{\mu}}^{\perp}\equiv\vec n.{\vec a}^{\perp}=0\ ,
\end{equation}
where $\vec n$ is a constant $p-$dimensional unit vector and we have suppressed the indices $i,j$. Next let us  consider any set of $N^2-N$ orthogonal matrices $R_{ij}\in SO(p)$ satisfying:
\begin{eqnarray}
&&R_{ij}.\vec n=\vec n_{ij}~~~{\rm for}~~~i>j;\\ 
 &&R_{ij}=R_{ji}~~~{\rm for}~~~i<j;  \nonumber\ .
\end{eqnarray}
Clearly such a set of matrices always exists. Now we define the Hermitian matrices:
\begin{equation} 
\vec a_{ij}\equiv R_{ij}.{\vec a_{ij}}^{\perp}~~~{\rm for}~~~i\neq j~~~{\rm and}~~~\vec a_{ii}\equiv 0; \ .
\end{equation}
The matrices $\vec a_{ij}$ are linearly dependent and satisfy the properties:
\begin{eqnarray}
&&\vec n_{ij}.\vec a_{ij}=0\\
&&\vec a_{ij}.\vec a_{ji}={\vec a_{ij}}^{\perp}.{\vec a_{ji}}^{\perp}
\end{eqnarray}
Next we define the change of coordinates:
\begin{equation}
X^{\mu}=U(x^{\mu}+a^{\mu})U^{-1};~~~U \in SU(N); \ . \label{coordinatechange}
\end{equation}
Note that on the left hand-side of equation (\ref{coordinatechange}) we have a set of $p$ $N\times N$ Hermitian matrices spanning a $pN^2$ dimensional linear space. On the other side by construction there are only $p-1$ linearly independent matrices labeled by $a^{\mu}$ and hence the dimension of the linear space spanned by  $x^{\mu}+a^{\mu}$ is equal to $(p-1)(N^2-N)+pN$. 
This suggests that in order to have a well defined change of coordinates  in equation (\ref{coordinatechange}) we need $N^2-N$ degrees of freedom, parameterizing the orbit of $x^{\mu}+a^{\mu}$ under the $SU(N)$ group. However a general element of $SU(N)$ has $N^2-1$ degrees of freedom. The $N-1$ degrees of freedom that are left out correspond to the stabilizer of the action of $SU(N)$ and as one can check are generated by a Cartan subalgebra of $su(N)$.

It is a straightforward exercise to compute the corresponding Jacobian. The expression that one obtains is given by (we refer the reader to Appendix~B for a detailed computation):
\begin{equation}
J=\left(\prod_{i\neq j}|\vec\Delta_{ij}|\right){\rm det}\Big|\Big|\delta_i^l\delta_j^m+Y_{ij}^{lm}\Big|\Big|{\rm det}\Big|\Big|\frac{\delta\theta_{rs}}{\delta u_{lm}}\Big|\Big|\ ,\label{Jacob}
\end{equation}
where $Y_{ij}^{lm}$ is given by:
\begin{equation}
Y_{ij}^{lm}=\frac{\vec n_{ij}}{|\vec\Delta_{ij}|}.(\vec a_{il}\delta_j^{m}-\vec a_{mj}\delta_i^{l})+\frac{\vec a_{ij}.\Pi_{ij}.\vec a_{ml}}{|\vec\Delta_{ij}|^2}.(\delta_i^{m}-\delta_i^{l}-\delta_j^{m}+\delta_j^{l})\ .
\end{equation}
Note that the last determinant in equation (\ref{Jacob}) is the Haar measure of $SU(N)$. Now we can write down the measure in the path integral (\ref{partfunct}) in terms of the new variables. To second order in the off-diagonal elements $\vec a_{ij}$ we have the expression:
\begin{equation}
\mu=DU\prod_i dx_i\left(\prod_{i\neq j}|\vec\Delta_{ij}|\right)\prod_{i>j}\left(d^{p-1}a_{ij}^{\perp}d^{p-1}a_{ji}^{\perp}\right)\left\{1-2\sum_{i\neq j}{\vec a}_{ij}.\hat D_{ij}.\vec a_{ji}+O(a^3)\right\}\ ,\label{measure1}
\end{equation}
where $DU$ is the Haar measure of $SU(N)$ and $\hat D_{ij}$ is given by:
\begin{equation}
D^{\mu\nu}_{ij}=\frac{\Pi_{ij}^{\mu\nu}}{|\vec\Delta_{ij}|^2}+\frac{1}{4}\sum_{l\neq i,j}\frac{n^{\mu}_{il}n^{\nu}_{jl}+n^{\nu}_{il}n^{\mu}_{jl}}{|\vec\Delta_{il}||\vec\Delta_{jl}|}\label{D}\ ,~~~\mbox{for}~~i \neq j\ .
\end{equation}

Note that without loss of generality one can take $\vec n=\vec {e}_p$. Note also that $\vec a_{ij}=R_{ij}.\vec a_{ij}^{\perp}$. Our next goal is to develop a systematic perturbative procedure to integrate out the off-diagonal degrees of freedom $\vec a_{ij}^{\perp}$.


\subsection{The effective action to two loops and stabilization }
In this subsection we will develop a perturbative technique to integrate out the off diagonal modes ${\vec a_{ij}}^{\perp}$ and calculate the semi-classical correction to the effective action for the diagonal modes $\vec x_i$.
 
Numerical studies of the three matrix model \cite{DelgadilloBlando:2012xg} give
a parabolic distribution of radius $R$ for the eigenvalues of one matrix, the unique
three dimensional rotationally invariant lift of which is a uniform
distribution inside a ball of the same radius, as argued in Appendix \ref{AppendixA2}.  The radius of
the distribution was found to be $R\approx 2.0>1$.  This suggests a
perturbative expansion in powers of $1/R$ may prove useful. In order
to proceed systematically we first rescale our variables $\vec x_i$
and $\vec a_{ij}$ in the following way:
 \begin{equation}
 \vec x=R{\tilde{\vec x}}_i;~~~\vec a=R{\tilde{\vec a}}_{ij}; \ .\label{rescale}
 \end{equation}
 Next we write the action (\ref{actionsplit}) as:
 \begin{equation}
S[{\tilde {\vec x}_i},{\tilde{\vec a}}_{ij}^{\perp}]=NR^4\sum_{i\neq j}\left(-\frac{1}{2}|\tilde{\vec\Delta}_{ij}|^2{\tilde{\vec a}}^{\perp}_{ij}.{\tilde{\vec a}}^{\perp}_{ji}+\tilde{\Delta}^{\mu}_{ij}{\tilde{a}}^{\nu}_{ji}[{\tilde{a}}^{\mu},{\tilde{a}}^{\nu}]_{ji}+\frac{1}{4}[{\tilde{a}}^{\mu},{\tilde{a}}^{\nu}]_{ij}[{\tilde{a}}^{\mu},{\tilde{a}}^{\nu}]_{ji}\right)\label{actiontilde}
 \end{equation}
 Note that from a field theory point of view the parameter $1/NR^4$ can be interpreted as a loop counting parameter. Next we define the correlation function:
 \begin{equation}
\langle{\cal O}\rangle_0=\frac{\int\prod_i d\tilde x_i\left(\prod_{i\neq j}|\tilde{\vec\Delta}_{ij}|\right)\prod_{i>j}\left(d^{p-1}{\tilde a}_{ij}^{\perp}d^{p-1}{\tilde a}_{ji}^{\perp}\right)\,{\cal O}\,e^{-NR^4\sum_{i\neq j}\frac{1}{2}|\tilde{\vec\Delta}_{ij}|^2{\tilde{\vec a}}^{\perp}_{ij}.{\tilde{\vec a}}^{\perp}_{ji}}}{\int\prod_i d\tilde x_i\left(\prod_{i\neq j}|\tilde{\vec\Delta}_{ij}|\right)\prod_{i>j}\left(d^{p-1}{\tilde a}_{ij}^{\perp}d^{p-1}{\tilde a}_{ji}^{\perp}\right)e^{-NR^4\sum_{i\neq j}\frac{1}{2}|\tilde{\vec\Delta}_{ij}|^2{\tilde{\vec a}}^{\perp}_{ij}.{\tilde{\vec a}}^{\perp}_{ji}}}\ .
\end{equation}
 For the propagator of ${\tilde{\vec a}}^{\perp}_{ij}$ we obtain:
 \begin{equation}
\langle{{\tilde a}^{\mu\perp}}_{ij}{\tilde{a}}^{\nu\perp}_{lm}
\rangle_0=\frac{1}{NR^4}\frac{(\delta^{\mu\nu}-n^{\mu}n^{\nu})}{|\tilde{\vec\Delta}_{ij}|^2}\delta_i^m\delta_j^l\ .\label{propperp}
 \end{equation}
In deriving (\ref{propperp}) one could use a frame in which $\vec n=\vec e_p$.
Note that the cubic and quartic contributions to the action (\ref{actiontilde}) as well as the Jacobian in the measure (\ref{measure1}) depend on ${\tilde{\vec a}}^{\perp}_{ij}$ through the relation  $\tilde{\vec a}_{ij}=R_{ij}.\tilde{\vec a}_{ij}^{\perp}$. This is why it is convenient to calculate the two-point function:
\begin{equation}
\langle {\tilde a}^{\mu}_{ij}{\tilde a}^{\nu}_{lm}\rangle_0=\frac{1}{NR^4}\frac{1}{|\tilde{\vec\Delta}_{ij}|^2}\Pi^{\mu\nu}_{ij}\delta_i^m\delta_j^l\ ,\label{prop}
\end{equation}
where we have used that:
\begin{equation}
R^{\mu\mu'}_{ij}.(\delta^{\mu'\nu'}-n^{\mu'}n^{\nu'}).R^{\nu\nu'}_{ij}=\delta^{\mu\nu}-n^{\mu}_{ij}n^{\nu}_{ij}=\Pi^{\mu\nu}_{ij}\ .
\end{equation}
We now have all the machinery required for a perturbative calculation in powers of $1/NR^4$. To leading order we obtain the following first loop effective action for the diagonal modes:
\begin{equation}
S^{(1)}_{\rm{eff}}(R,\tilde {\vec x})=[(p-2)N^2-2(p-1)N+p]\log R+\frac{(p-2)}{2}\sum_{i\neq j}\log |{\tilde{\vec\Delta}}_{ij}|^2\ .\label{firstloop}
\end{equation}  
As one can see from the first term in equation (\ref{firstloop}) at large $N$ 
and for $p\geq 2$ at one loop the effective action gives an attractive potential $V_{\rm{eff}}(R)$ and is therefore not sufficient to stabilize the radius of 
the distribution. On the other hand the second loop corrections has 
an overall factor of  $1/NR^4$ and
could balance the $\log R$ attractive potential in (\ref{firstloop}). 

At second loop the effective action has contributions from the cubic 
and quartic vertices in (\ref{actiontilde}) as well as from the quadratic 
term in the measure (\ref{measure1}) (``ghost's contribution"). The corresponding correlation functions are: $\langle(NR^4\tilde{\Delta}^{\mu}_{ij}{\tilde{a}}^{\nu}_{ji}[{\tilde{a}}^{\mu},{\tilde{a}}^{\nu}]_{ji})^2\rangle_0$, $\langle NR^4\frac{1}{4}tr[{\tilde{a}}^{\mu},{\tilde{a}}^{\nu}]^2\rangle_0$ and $\langle -2\sum_{i,j}{\tilde{\vec a}}_{ij}.D_{ij}.\tilde{\vec a}_{ji}\rangle_0$. Using Wick contractions and the two-point function (\ref{prop}) we can calculate the second loop contribution. After somewhat tedious but straightforward calculations we obtain:
\begin{eqnarray}
\langle(NR^4\,\sum_{i\neq j}\,\tilde{\Delta}^{\mu}_{ij}{\tilde{a}}^{\nu}_{ji}[{\tilde{a}}^{\mu},{\tilde{a}}^{\nu}]_{ji})^2\rangle_0&=&\frac{1}{2NR^4}\sum_{i,j\neq l}\frac{\{(4p-6)\sin^2\theta_{i,jl}+\sin^2\theta_{l,ij}\}}{\tilde{\vec \Delta}_{il}^2\tilde{\vec \Delta}_{jl}^2}\ ,\\
\langle NR^4\frac{1}{4}tr[{\tilde{a}}^{\mu},{\tilde{a}}^{\nu}]^2\rangle_0&=&\frac{-1}{2NR^4}\sum_{i,j\neq l}\frac{\{(p-1)(p-2)+\sin^2\theta_{l,ij}\}}{\tilde{\vec \Delta}_{il}^2\tilde{\vec \Delta}_{jl}^2}\ ,\\
\langle -2\sum_{i\neq j}{\tilde{\vec a}}_{ij}.D_{ij}.\tilde{\vec a}_{ji}\rangle_0&=&\frac{-1}{2NR^4}\sum_{i,j\neq l}\frac{2\sin^2\theta_{i,jl}}{\tilde{\vec \Delta}_{il}^2\tilde{\vec \Delta}_{jl}^2}+O\left(\frac{1}{N}\right)\ .\label{ghosts}
\end{eqnarray}
Were the angles $\theta_{l,ij}, \theta_{i,jl}$ are defined via $\cos\theta_{i,jl}=\vec n_{ij}.\vec n_{il}$ and the last term in equation (\ref{ghosts}) corresponds to non-planar diagram contributions, subleading in the large $N$ limit. For the total second loop contribution to the effective potential we obtain:
\begin{equation}
S_{\rm{eff}}^{(2)}(R,\tilde{\vec x})=\frac{(p-2)}{2NR^4}\sum_{i,j\neq l}\frac{p-1-4\sin^2\theta_{i,jl}}{\tilde{\vec \Delta}_{il}^2\tilde{\vec \Delta}_{jl}^2}\ .\label{secondlooppot}
\end{equation}
The full large $N$ effective action for the one loop is then
\begin{equation}
S_{\rm{eff}}(R,\tilde{\vec x})=\frac{(p-2)}{2}\left(\sum_{i\neq j}\ln (R^2\Delta_{i,j}^2)+\frac{1}{NR^4}\sum_{i,j\neq l}\frac{p-1-4\sin^2\theta_{i,j,l}}{\tilde{\vec \Delta}_{il}^2\tilde{\vec \Delta}_{jl}^2}\right)\ .\label{effectiveaction}
\end{equation}
In the large $N$ limit we can trade the discrete sums in 
equations (\ref{firstloop}),(\ref{secondlooppot}) for integrals 
over a joint distribution $\rho(\tilde{\vec x})$ via:
\begin{equation}
\frac{1}{N}\sum_i\rightarrow\int_{B^p} d^p\tilde x\rho(\tilde{\vec x}); \ ,
\end{equation}
where the integral is over a ball $B^p$ of unit radius. In the large
$N$ limit we obtain the following second loop effective potential for
the radius, $R$, of the distribution:
\begin{equation}
V_{\rm{eff}}(R)=(p-2)N^2\left(\log R+\frac{\#(p)}{R^4}\right)\ ,\label{effpotR}
\end{equation}
where we have defined:
\begin{equation}
\#(p)=\frac{1}{2}\int_{B^p}\int_{B^p}\int_{B^p}d^p\tilde x d^p\tilde yd^p\tilde z \rho(\tilde{\vec x}) \rho(\tilde{\vec y}) \rho(\tilde{\vec z})\frac{p-1-4\sin^2\theta_{x,yz}}{(\tilde{\vec x}-\tilde{\vec z} )^2(\tilde{\vec y}-\tilde{\vec z} )^2}\label{numberp}
\end{equation}
and $\theta_{x,yz}$ is the analog of $\theta_{i,jl}$ defined via:
\begin{equation}
\cos\theta_{x,yz}=\frac{(\tilde{\vec x}-\tilde{\vec y} ).(\tilde{\vec x}-\tilde{\vec z} )}{|(\tilde{\vec x}-\tilde{\vec y} )||(\tilde{\vec x}-\tilde{\vec z} )|}\ .\label{cos}
\end{equation}
We can now estimate the extent of the distribution by minimizing (\ref{effpotR}). We obtain:
\begin{equation}
R_p=(4\#(p))^{1/4};\ .\label{R}
\end{equation}
Note that for $p\geq 5$ we have $\#(p)>0$ (the integrand in equation
(\ref{numberp}) is non-negative) and the ground state stabilizes at
the finite radius estimated in equation (\ref{R}). For $p=3$ numerical
simulations \cite{DelgadilloBlando:2012xg} give excellent agreement
with a parabolic distribution for the eigenvalues of a single matrix.
As shown in Appendix \ref{AppendixA2} the unique lift to a
rotationally invariant distribution gives the uniform distribution. We
therefore evaluated analytically the integral in equation
(\ref{numberp}), assuming a uniform distribution
$\rho(\tilde{\vec x})=\rm{const}$ (see Appendix~C for more
details). The resulting radius is:
\begin{equation}
R_3=\left(9-\frac{3}{5}\pi^2\right)^{1/4}\approx 1.323; \ .
\end{equation}
Equations (\ref{effpotR})-(\ref{R}) contain the main result of our 
perturbative calculation. In the next section we discuss the gauge 
dependence of our expression for the radius of the distribution $R_p$ in equation (\ref{R}).


\subsection{Gauge dependence}
In subsection 3.1 we outlined a change of coordinates that was equivalent to introducing the gauge $\vec n_{ij}.\vec a_{ij}=0$. Alternatively we could have used standard Faddeev-Popov techniques to fix our gauge. Let us consider the gauge condition:
\begin{equation}
f_{ij}=\vec\Delta_{ij}.\vec a_{ij}=0; \label{gaugeFP}\ .
\end{equation}
The corresponding Faddeev-Popov determinant is given by:
\begin{equation}
\Delta_{FP}=\prod_{i>j}|\vec\Delta_{ij}|^2{\rm det}||\delta_{i}^l\delta_j^m+{Y'}_{ij}^{lm}|| \ ,
\end{equation}
where:
\begin{equation}
{Y'}_{ij}^{lm}=\frac{\vec n_{ij}}{|\vec\Delta_{ij}|}.(\vec a_{il}\delta_j^{m}-\vec a_{mj}\delta_i^{l})+\frac{\vec a_{ij}.\vec a_{ml}}{|\vec\Delta_{ij}|^2}.(\delta_i^{m}-\delta_i^{l}-\delta_j^{m}+\delta_j^{l})\ .\label{FPdet}
\end{equation}
Notice that the gauge condition $f^{\eta}_{ij}=\vec\Delta_{ij}.\vec a_{ij}-\eta_{ij}=0$ would result to the same Faddeev-Popov determinant (\ref{FPdet}). Now integrating over the family of gauge functions $f^{\eta}_{ij}$ with weight $\exp(-|\eta_{ij}|^2/2\xi)$ would modify the action (\ref{actionsplit}) to:
\begin{equation}
S[X,\xi]=N\frac{1}{2}\sum_{i\neq j} |\vec\Delta_{ij}|^2{a}^{\mu}_{ij}(\Pi^{\mu\nu}_{i j}+\frac{1}{\xi}{n^{\mu}_{ij}n^{\nu}_{ij}})a^{\nu}_{ji}-N\sum_{i\neq j}\Delta^{\mu}_{ij} a^{\nu}_{ij}[a^{\mu},a^{\nu}]_{ji}-N\frac{1}{4}tr[a^{\mu},a^{\nu}]^2 \ ,\label{actionXi}
\end{equation}
Next we can go through the steps considered in section 3.2, namely rescale with the radius of the joint eigenvalue distribution $R$ as in equation (\ref{rescale}) and set up perturbative calculation in powers of $1/R$. One can show that the first loop effective action $V^{(1)}_{\rm eff}$ is still given by equation (\ref{firstloop}) and is thus gauge independent. However the two-point function (\ref{prop}) (the propagator for $\tilde{\vec a}_{ij})$ is modified to:
\begin{equation}
\langle {\tilde a}^{\mu}_{ij}{\tilde a}^{\nu}_{lm}\rangle_{\xi}=\frac{1}{NR^4}\frac{1}{|\tilde{\vec\Delta}_{ij}|^2}(\Pi^{\mu\nu}_{ij}+\xi n^{\mu}_{ij}n^{\nu}_{ij})\delta_i^m\delta_j^l\ .\label{propXi}
\end{equation}
Note that the result from equation (\ref{prop}) corresponds to the choice $\xi=0$ (Landau gauge). Let us consider the choice $\xi=1$ (Feynman gauge) and calculate the second loop contribution to the effective action. Going through the same steps as in section~3.2 we obtain the analog of equation (\ref{effpotR}):
\begin{equation}
V_{\rm{eff}}(R)=(p-2)N^2\left(\log R+\frac{\tilde\#(p)}{R^4}\right)\ ,\label{effpotRFn}
\end{equation}
where $\tilde\#(p)$ is given by:
\begin{equation}
\tilde\#(p)=\frac{p-2}{2}\int_{B^p}\int_{B^p}\int_{B^p}d^p\tilde x d^p\tilde yd^p\tilde z \rho(\tilde{\vec x}) \rho(\tilde{\vec y}) \rho(\tilde{\vec z})\frac{1}{(\tilde{\vec x}-\tilde{\vec z} )^2(\tilde{\vec y}-\tilde{\vec z} )^2}\label{numberp'}
\end{equation}
This results in the radius:
\begin{equation}
R'_p=(4\tilde\#(p))^{1/4};\ .\label{R'}
\end{equation}
It is clear from equation (\ref{numberp'}) that $\tilde\#(p)$ is positive for $p\geq3$. For $p=3$ we have evaluated analytically $\tilde\#(3)$ (look at Appendix C). The corresponding radius is:
\begin{equation}
R'_3=\left(\frac{9}{2}+\frac{3}{5}\pi^2\right)^{1/4}\approx 1.797\ .
\end{equation}

Apparently the results obtained in Landau and Feynman gauges differ. In order to address the issue of gauge dependence let us focus on a particular representative of the family of gauge conditions $f^{\eta}_{ij}=\vec\Delta_{ij}.\vec a_{ij}-\eta_{ij}=0$. Note that the change of coordinates (\ref{coordinatechange}) considered in section 3.2 implements the $\eta_{ij}=0$ case. One can show that the gauge condition for general $\eta_{ij}$ can be implemented along the lines of section 3.2 via the following modified change of coordinates:
\begin{equation}
X^{\mu}=U||x_i^{\mu}\delta_{ij}+R^{\mu\nu}_{ij}a^{\nu\perp}_{ij}+\frac{n^{\mu}_{ij}}{|\vec\Delta_{ij}|}\eta_{ij}||U^{-1};~~~U \in SU(N); \ . \label{coordinatechangeETA}
\end{equation}
Let us suppose that the theory has settled in its ground state which is a commuting phase. There should exist unitary matrix $V\in SU(N)$ which simultaneously diagonalizes the $X^{\mu}$ matrices and hence we can write:
\begin{equation} 
V^{-1}\lambda^{\mu}V=U||x_i^{\mu}\delta_{ij}+R^{\mu\nu}_{ij}a^{\nu\perp}_{ij}+\frac{n^{\mu}_{ij}}{|\vec\Delta_{ij}|}\eta_{ij}||U^{-1}\ , \label{diagonalized}
\end{equation}
where $\lambda^{\mu}$ is a diagonal matrix. Now if we square equation (\ref{diagonalized}), take a trace over the gauge indices and sum over $\mu$ we obtain:
\begin{equation}
\sum_{i}\vec\lambda_i^2=\sum_{i}\vec x_i^2+\sum_{i\neq j}|{{\vec a}^{\perp}_{ij}}|^2+\sum_{i\neq j}\frac{|\eta_{ij}|^2}{|\vec\Delta_{ij}|^2}\geq\sum_{i}\vec x_i^2+\sum_{i\neq j}\frac{|\eta_{ij}|^2}{|\vec\Delta_{ij}|^2}\ .
\end{equation}
Next we define average radii of the distribution $r_{\lambda}$ and $r_x$ via:
\begin{equation}
r_{\lambda}^2=\frac{1}{N}\sum_{i}\vec\lambda_i^2;~~~r_x^2=\frac{1}{N}\sum_{i}\vec x_i^2;
\end{equation}
and learn that:
\begin{equation}
r_{\lambda}^2-r_x^2\geq \frac{1}{N}\sum_{i\neq j}\frac{|\eta_{ij}|^2}{|\vec\Delta_{ij}|^2}\ .
\end{equation}
Therefore the average radius of the eigenvalue distribution
$r_{\lambda}$ always differs from the average radius of the
distribution of the diagonal modes $r_x$, unless $\eta_{ij}=0$ or the
eigenvalues are infinitely spread in which case there is no well
defined average radius. This could explain why the gauge fixing 
procedure outlined
above, which involved averaging over all possible values of
$\eta_{ij}$ failed to produce a gauge independent answer for the
radius of the eigenvalue distribution.
These consideration suggests that the
gauge $\eta_{ij}=0$  should be optimal for describing the almost 
commuting theory.  

Alternatively one could take the point of view
that both gauge choices are equally valid and describe different
approximations to the true result, they only differ due to the intrinsic 
errors in a perturbative calculation. If we take this point of view
we can use the difference to estimate the errors in our estimate of $R$.
If we do this we conclude that $R\sim1.6\pm0.5$ which is in reasonable agreement
with the numerical results.



\section{Discussion}

In this paper we have followed two threads, in the first we investigated 
the $3$-matrix model of \cite{Hoppe:PhDThesis1982,Kazakov:1998ji} and find
that in the large $g$ limit the $3$-matrices commute and have a joint
eigenvalue distribution given by the uniform distribution within a ball of 
radius $R={\left(\frac{3\pi}{2g}\right)}^{1/3}$ . We show that a simple effective potential for the radius of the distribution reproduces the exact result.
Furthermore, in Appendix A we first demonstrated how to do perturbation theory around the uniform distribution and then how this distribution is the unique 
rotational invariant lift of the eigenvalue distribution of a single matrix.

Encouraged by the success of this effective potential calculation 
we develop an effective potential for the radius of 
the $p$-component Yang-Mills matrix model to two loops. 
We have done this by deriving an effective potential for the diagonal modes 
while preserving $SO(3)$ invariance and then assuming that these modes 
are uniformly distributed. The direct analog of the computations
in earlier sections would be a two loop computation in 
the axial gauge (where one of the matrices is diagonalized). Unfortunately
this gauge choice leads to infrared divergences at two loops and 
so we have not pursued this option.

We found that it is necessary to go to two loops as at one loop 
the effective potential is not stable since there is no classical 
potential and the one loop term gives
an attractive potential which is a rotationally invariant version 
of the Vandermonde determinant. The eigenvalue repulsion arises 
at two loops and gives a $\frac{1}{R^4}$ hard core potential. It is easy to see
that higher order terms give inverse higher powers of $R$ and our two loop 
estimate can only be a very rough approximation. 

Our estimate for the radius is unfortunately gauge dependent with $R=1.323$ in the Landau gauge and $R=1.797$ in the Feynman gauge. It is reasonable to assume
that the difference between these is an indication of the errors in the method
which would indicate that perhaps a reasonable estimate can be obtained
by averaging the two and taking the difference as an indication of the error yielding the prediction $R=1.6\pm 0.5$.  An alternative approach pursued by
Hotta, Nishimura and Tsuchiya \cite{Hotta:1998en} examined similar 
questions, for some observables, in the general Yang-Mills $p$-matrix model\footnote{Hotta et al 
\cite{Hotta:1998en} also obtained a two loop 
effective action for eigenvalues of the matrices $X_\mu$ by integrating 
out the $U(N)$ transformations that diagonalize the matrices. 
This yields a non-rotationally invariant effective action, which does not 
lead to a stable effective potential for the radial extent of the eigenvalues.}.

If we take their result 
\begin{equation}
<\frac{\rm{tr}}{N}(X_a^2)>=\sqrt{\frac{p}{2}}(1+\frac{7}{6p}+\cdots)
\label{largepexpansion}
\end{equation}
and assume that this is valid for $p=3$ together with the assumption 
that the eigenvalue distribution of a single matrix is parabolic of extent $R$ (which is consistent with a uniform joint distribution for commuting matrices) 
then $<\frac{\rm{tr}}{N}(X_a^2)>=3\frac{R^2}{5}$ and 
using (\ref{largepexpansion}) we obtain the estimate $R=1.68$, which is surprisingly close to the estimate we obtain above. In this case 
one can also attempt an estimate of the error by noting that if 
instead of (\ref{largepexpansion}) we use 
$<\frac{\rm{tr}}{N}(X_a^2)>=
\sqrt{\frac{p}{2}}(\frac{1}{(1-\frac{7}{6p})}+\cdots)$
which has the same leading large $p$ expansion we obtain $R=1.83$ and
therefore estimate the error (within the assumption of a parabolic 
distribution) that $R=1.75\pm.15$ 
which is in surprisingly good agreement with Monte Carlo 
simulations \cite{DelgadilloBlando:2012xg} which give a value for $R=2.0$

We conclude that, though the random matrices are not commuting, 
a useful approximation is to take the background formed by these 
fluctuating matrices as that of commuting matrices whose 
joint eigenvalue distribution is approximately uniform within a ball of radius 
$R$. This background gives reasonable agreement with the
numerical work and serves as a reasonable starting point for further
work.


\section{Acknowledgments}

The work of V. F. was funded by an INSPIRE IRCSET-Marie Curie International Mobility Fellowship.

\appendix

\section {Corrections at large coupling and lifting the distribution}

In this appendix we provide a more rigorous derivation of the uniform
distribution obtained in section 2.2. We first develop perturbation
theory around the uniform background and then obtain the leading
correction to the uniform distribution for large coupling. 

In the second section we show that the uniform distribution is the unique
rotationally invariant lift of the parabolic distribution and more
generally establish the rotational invariant lifts of a one
dimensional distribution to three and then down to two dimensions.
\subsection{Iterative Solution}
\label{AppendixA1}
Here we derive an integral equation of the second kind for the correction to the uniform distribution (\ref{rho}). Our starting point is the integral equation (\ref{IE1}):
\begin{equation}
1=\int d^3x'\,\frac{\rho(\vec x')}{|\vec x-\vec x'|^2(1+g^2|\vec x-\vec x' |^2)}\ ,~~~~\mbox{for}~ |\vec x|\leq R\  ,  \label{A1}
\end{equation}
where the integral is performed over a ball of radius $R$. Since the radius is $g$ dependent (with leading order dependence given by equation (\ref{radius}) ) it is more convenient to introduce new variables:
\begin{equation}
 \vec\eta =\vec x/R\ ,~~~ \tilde\rho(\vec\eta)=R\,\rho(R\,\vec\eta)\ . \label{A1'}
\end{equation}
 In these variables the integral equation (\ref{A1}) becomes:
\begin{equation}
1=\int d^3\eta'\,\frac{\tilde\rho(\vec \eta')}{|\vec \eta-\vec \eta'|^2(1+(Rg)^2|\vec \eta-\vec \eta' |^2)}\ ,~~~~\mbox{for}~ |\vec \eta|\leq 1\  ,  \label{A2}
\end{equation}
where the integral is performed over a ball of unit radius. Note that from equation (\ref{radius}) it follows that at large $g$, $Rg\sim g^{2/3}$ and $Rg$ is also large. Therefore we are interested in solving equation (\ref{A2}) at large $Rg$. The fact that the kernel of equation (\ref{A2}) is a delta convergent series:
\begin{equation}
\frac{1}{|\vec \eta-\vec \eta'|^2(1+(Rg)^2|\vec \eta-\vec \eta' |^2)}=\frac{2\pi^2}{Rg}\delta(\vec\eta-\vec\eta') + \dots \ ,
\end{equation}
suggests that we make the following definitions:
\begin{equation}
\Delta K(\vec\eta-\vec\eta',Rg) =\frac{Rg}{2\pi^2|\vec \eta-\vec \eta'|^2(1+(Rg)^2|\vec \eta-\vec \eta' |^2)}-\delta(\vec\eta-\vec\eta')\ ,~~~\Delta\tilde\rho(\vec\eta)=\tilde\rho(\vec\eta)-\frac{Rg}{2\pi^2}\ .
\end{equation}
Note that $Rg/(2\pi^2)$ corresponds to the uniform distribution (\ref{rho}) in the variables (\ref{A1'}), which means that $\Delta\rho$ is the correction to the uniform distribution in the same variables. It is straightforward to write the integral equation (\ref{A2}) in terms of $\Delta K$ and $\Delta\tilde\rho$:
\begin{equation}
\Delta\tilde\rho(\eta)=-\frac{Rg}{2\pi^2}\int\,d^3\,\eta' \,\Delta K(\vec\eta-\vec\eta',Rg)-\int\,d^3\eta' \,\Delta K(\vec\eta-\vec\eta',Rg)\,\Delta\tilde\rho(\vec\eta')\ .\label{A6}
\end{equation}
Equation (\ref{A6}) is an integral equation of the second kind for the function $\Delta\tilde\rho$ and can be developed recursively as the series:
\begin{equation}
\Delta\tilde\rho=-\frac{Rg}{2\pi^2}\,\left[ \int\Delta K -\int\Delta K\int\Delta K +\int\Delta K\int\Delta K\int\Delta K- \dots   \right]\ , \label{A7}
\end{equation}
where we used condensed notations. One can check that $\int d^3\eta' \Delta K \sim 1/(Rg)$ for $|\vec\eta|<1$, which suggests that the series is convergent for large $Rg$. This can be verified numerically. If the series is convergent this is sufficient for the integral equation (\ref{A6}) to have a unique solution given by (\ref{A7}).

Here we present the first iteration (the leading order) correction to the uniform distribution defined by:
\begin{equation}
\tilde\rho_{(1)}(\eta,Rg)=-\frac{Rg}{2\pi^2}\int\,d^3\,\eta' \,\Delta K(\vec\eta-\vec\eta')\ .
\end{equation}
Our result is:

\begin{eqnarray}
\tilde\rho_{(1)}(Rg,\eta)&=&\frac{Rg}{2\pi^3}\left[\pi-\tan^{-1}[Rg(1+\eta)]-\tan^{-1}[Rg(1-\eta)]\right]-\frac{1}{8\pi^3\,\eta}\log\left[\frac{1+(Rg)^2(1-\eta)^2}{1+(Rg)^2(1+\eta)^2} \right]\nonumber\\
&-&\frac{(Rg)^2}{8\pi^3\,\eta}\,(1-\eta^2)\,\log\left[\frac{(1+\eta)^2(1+(Rg)^2(1-\eta)^2)}{(1-\eta)^2(1+(Rg)^2(1+\eta)^2)}\right]\ .\label{A8}
\end{eqnarray}
Note that $\tilde\rho_{1}(\eta,Rg)$ is regular and of order one for $\eta\in[0,1)$. One can check that $\int d^3\,\eta\,\tilde\rho_{(1)}(\eta,Rg) \sim \log(Rg)$, which is subleading relative to the uniform distribution $Rg/(2\pi^2)$ whose integral is of order $Rg$. 

We can perform a non-trivial check of this results by calculating the observable:
\begin{equation}
\nu=\frac{1}{3}g^2\,\int d^3 x \rho(\vec x){\vec x}^2 =\frac{1}{3}R^2(Rg)^2\int d^3\eta\,\tilde\rho(\vec\eta)\,{\vec\eta}^2\ , \label{A9}
\end{equation}
obtained in closed form in ref.~\cite{Kazakov:1998ji}. Note that the radius $R$ can be extracted from $\tilde\rho(\eta)$ through the relation:
\begin{equation}
\int d^3\eta\tilde\rho(\eta) =R^{-2}\ . \label{A10}
\end{equation}
Using the leading order result for $\tilde\rho$:
\begin{equation}
\tilde\rho(\eta,Rg)=\frac{Rg}{2\pi^2}+\tilde\rho_{(1)}(\eta,Rg)+O((Rg)^{-1})\ ,
\end{equation}
it is straightforward to obtain:
\begin{eqnarray}
&&R^{-2}=\frac{2Rg}{3\pi}+\frac{2\log(2Rg)+1}{2\pi^2}+\dots \ , \label{A13}\\
&&\int d^3\eta\,\tilde\rho(\vec\eta)\,{\vec\eta}^2=\frac{2Rg}{5\pi}+\frac{2\log(2Rg)-1}{2\pi^2}+\dots \ ,\label{A14}
\end{eqnarray}
substituting equations (\ref{A13}) and (\ref{A14}) in equation (\ref{A9}), we can obtain large $Rg$ expansion of $\nu$. Furthermore from equation (\ref{A14}) we can obtain the large $g$ expansion of $Rg$. Therefore, we can calculate the first two leading terms in the large $g$ expansion of $\nu$:
\begin{equation}
\nu=\frac{(12\pi)^{2/3}}{20}g^{4/3}-\frac{(12\pi)^{1/3}}{4\pi}g^{2/3}+\dots\ ,\label{A15}
\end{equation} 
which is exactly the result of ref.~\cite{Kazakov:1998ji}.


\subsection{``Lifting" the 1d distribution}
\label{AppendixA2}
In this subsection we derive a relation between the two and three
dimensional distributions considered in section 2.2 and the one
dimensional distribution considered in section 2.1 We begin by
discussing the lift to the three dimensional distribution.

Let us consider
equation (\ref{EQN3}), if we apply the operator $\vec n.\nabla_{x}$ on
both sides of the equation we obtain:
\begin{equation}
\vec n.\vec x=\int \frac{d^3x'\rho(\vec x')}{(\vec n.(\vec x-\vec x'))(1+g^2(\vec n.(\vec x-\vec x'))^2)}\label{A16}
\end{equation}
Next if we define $u=\vec n.\vec x$, choose a coordinate frame in the integral along $\vec x'$ in which we have $\vec n.\vec x'=x'^1$ and define:
\begin{equation}
\rho_1(x_1)=\int\limits_{-\sqrt{R^2-x_1^2}}^{\sqrt{R^2-x_1^2}}dx_2\int\limits_{-\sqrt{R^2-x_1^2-x_2^2}}^{\sqrt{R^2-x_1^2-x_2^2}}dx_3\,\rho(\vec x)\ ,\label{A17}
\end{equation}
we obtain equation (\ref{Eqn2}):
\begin{equation}
u=\int \frac{dx'^1\rho_1( x'^1)}{(u-x'^1)(1+g^2(u-x'^1)^2)}\label{AEqn2}\ ,
\end{equation}
which we solved in section 2.1 at large $g$ to obtain the parabolic distribution (\ref{parabolic}) of ref.~\cite{Berenstein:2008eg}. To find the corresponding three dimensional (``lifted") distribution we have to solve the integral equation (\ref{A17}). This can easily be achieved by integrating over the disc $x_2^2+x_3^2\leq R^2-x_1^2$ in cylindrical coordinates:
\begin{equation}
\rho_1(x_1)=\int\limits_{0}^{\sqrt{R^2-x_1^2}}dr\,(2\pi r)\,\rho(\sqrt{x_1^2+r^2})=\int\limits_{x_1}^{R}dw\,(2\pi w)\rho(w)\ , \label{A18}
\end{equation}
where we substituted $r=\sqrt{w^2-x_1^2}$. By differentiating\footnote{Note that there are no subtleties in differentiating near the boundary since $\rho_1({R})=0$.} equation (\ref{A18}) with respect to $x_1$ we can solve for $\rho$ in therms of $\rho_1$:
\begin{equation}
\rho(x)=-\frac{\rho_1'(x)}{2\pi\,x}\ . \label{A19}
\end{equation}
Equation (\ref{A19}) is the unique solution for a ``lifted" rotationally invariant three dimensional distribution reproducing the one dimensional distribution $\rho_1$. Furthermore since at large $g$ the three matrix model is commuting this suggests that the lifted distribution is the joint eigenvalue distribution of the model. It is easy to check that if one substitutes the parabolic distribution (\ref{parabolic}) with the radius (\ref{radL}) into equation (\ref{A19}) one obtains exactly the uniform distribution (\ref{rho}):
\begin{equation}
\rho(\vec x)=\frac{g}{2\pi^2}\ .
\end{equation}
We can now reduce to the two dimensional distribution by integrating over the 
additional coordinate.
\begin{equation}
\rho_2(\vec{r})=-\int_{-\sqrt{R^2-r^2}}^{\sqrt{R^2-r^2}}\frac{\rho_1'(\sqrt{r^2+z^2)}}{2\pi\,\sqrt{r^2+z^2}}dz\ , \label{A20}
\end{equation}
which in the large $g$ limit reproduces (\ref{hemi-sphere}). 

\section{The Jacobian}
In this section we outline the calculation of the Jacobian (\ref{Jacob}). Let us begin by differentiating equation (\ref{coordinatechange}). We obtain:
\begin{equation}
(U^{-1}d\vec X U)_{ij}=d\vec x_i\delta_{ij}+dR_{ij}.\vec a^{\perp}_{ij}+R_{ij}.d\vec a^{\perp}_{ij}-|\vec\Delta_{ij}|\vec n_{ij}\theta_{ij}-[a,\theta]_{ij}\ ,
\end{equation} 
where we have defined the Maurer-Cartan form $\theta\equiv U^{-1}dU$. Next we define the tetrads $\vec E$ via:
\begin{equation}
\vec E_i=(U^{-1}d\vec X U)_{ii};~~~\vec E_{ij}=(U^{-1}d\vec X U)_{ij}~~~{\rm for}~~~i\neq j; \ .
\end{equation}
In matrix form we have:

\begin{table}[h]
{\Large
$||E||$= {\begin{tabular}{|c|c|c|c|} \hline
 &$d\vec x_k$&$R_{rs}.d\vec a^{\perp}_{rs}$&$du_{lm}$\\ \hline
 &&&\\
$\vec E_i$&$\hat 1\delta_{i}^k$&0&$-\frac{\delta[\vec a,\theta]_{ii}}{\delta u_{lm}}$\\
 &&&\\ \hline
  &&&\\ 
$\vec E^{\perp}_{ij}$&$\Pi_{ij}.\frac{\partial R_{ij}}{\partial \vec x_k}.\vec a^{\perp}_{ij}$&$\hat 1\delta_{i}^r\delta_j^s$&$-\Pi_{ij}.\frac{\delta[\vec a,\theta]_{ij}}{\delta u_{lm}}$ \\
 &&&\\ \hline
  &&&\\ 
$E_{ij}^{||}$&$\frac{\partial \vec n_{ij}}{\partial \vec x_k}.\vec a_{ij}$&0&$|\vec\Delta_{ij}|\frac{\delta\theta_{ij}}{\delta u_{lm}}+\vec n_{ij}.\frac{\delta[\vec a,\theta]_{ij}}{\delta u_{lm}}$\\
 &&&\\ \hline
\end{tabular}}}\ ,
\end{table}

where $\Pi_{ij}=\hat 1-\vec n_{ij}\vec n_{ij}$ and we have split the off-diagonal tetrads $\vec E_{ij}$ into:
\begin{equation} 
\vec E_{ij}^{\perp}=\Pi_{ij}.\vec E_{ij};~~~{\rm and}~~~E_{ij}^{||}=-\vec n_{ij}.\vec E_{ij}; \ .
\end{equation}
We have also used that:
\begin{equation}
\vec n_{ij}.\frac{\partial R_{ij}}{\partial x_k}+\frac{\partial \vec n_{ij}}{\partial x_k}.R_{ij}=0\ .
\end{equation}
Now the Jacobian of interest, $J$, is given by the determinant of $||E||$. It is an easy exercise to show that the Jacobian is given by:
\begin{equation}
J={\rm det}\Big|\Big| |\vec\Delta_{ij}|\frac{\delta\theta_{ij}}{\delta u_{lm}}+\vec n_{ij}.\frac{\delta[\vec a,\theta]_{ij}}{\delta u_{lm}}+\sum_k\frac{\partial \vec n_{ij}}{\partial x_k}.\vec a_{ij}\frac{\delta[\vec a,\theta]_{kk}}{\delta u_{lm}}\Big|\Big|\label{Jac}
\end{equation}
One can show that the determinant in equation (\ref{Jac}) factorizes:
\begin{equation}
J={\rm det}\Big|\Big| |\vec\Delta_{ij}|\delta_i^r\delta_j^s+\vec n_{ij}.\frac{\delta[\vec a,\theta]_{ij}}{\delta\theta_{rs}}+\sum_{\mu,k}\frac{\partial \vec n_{ij}}{\partial x^{\mu}_k}.\vec a_{ij}\frac{\delta[a^{\mu},\theta]_{kk}}{\delta\theta_{rs}}\Big|\Big|{\rm det}\Big|\Big|\frac{\delta\theta_{rs}}{\delta u_{lm}}\Big|\Big|\label{Jfact}
\end{equation}
The last determinant in equation (\ref{Jfact}) produces the correct measure of $SU(N)$ and it is the first part $J'$ that we are really interested in. One can verify that:
\begin{equation}
J'=\left(\prod_{i\neq j}|\vec\Delta_{ij}|\right){\rm det}\Big|\Big|\delta_i^l\delta_j^m+Y_{ij}^{lm}\Big|\Big|\ ,
\end{equation}
where $Y_{ij}^{lm}$ is given by:
\begin{equation}
Y_{ij}^{lm}=\frac{\vec n_{ij}}{|\vec\Delta_{ij}|}.(\vec a_{il}\delta_j^{m}-\vec a_{mj}\delta_i^{l})+\frac{\vec a_{ij}.\Pi_{ij}.\vec a_{ml}}{|\vec\Delta_{ij}|^2}.(\delta_i^{m}-\delta_i^{l}-\delta_j^{m}+\delta_j^{l})\ ,
\end{equation}
where we have used that:
\begin{equation}
\frac{\delta[\vec a,\theta]_{ij}}{\delta\theta_{lm}}=(\vec a_{il}\delta_j^m-\vec a_{mj}\delta_i^l);~~~\frac{\partial n_{ij}^{\mu}}{\partial x^{\nu}_k}=\frac{\Pi^{\mu\nu}_{ij}}{|\vec\Delta_{ij}|}(\delta_i^k-\delta_j^k);~~~\vec a_{ij}.\Pi_{ij}=\vec a_{ij}; \ .
\end{equation}
Now using that:
\begin{equation}
{\rm det}\Big|\Big|\delta_i^l\delta_j^m+Y_{ij}^{lm}\Big|\Big|={\rm exp}\left\{{\rm tr}Y-\frac{1}{2}{\rm tr}Y^2+O(Y^3)\right\}
\end{equation}
One can easily verify that:
\begin{equation}
J'={\rm exp}\left\{\sum_{i\neq j}\log|\vec\Delta_{ij}|-2\sum_{i\neq j}{\bar{\vec a}}_{ij}.\hat D_{ij}.\vec a_{ji}+O(a^3)\right\}\ ,
\end{equation}
where $\hat D_{ij}$ is given by:
\begin{equation}
D^{\mu\nu}_{ij}=\frac{\Pi_{ij}^{\mu\nu}}{|\vec\Delta_{ij}|^2}+\frac{1}{4}\sum_{k}\frac{n^{\mu}_{ik}n^{\nu}_{jk}+n^{\nu}_{ik}n^{\mu}_{jk}}{|\vec\Delta_{ik}||\vec\Delta_{jk}|}\label{D}\ .
\end{equation}
Therefore our final expression for the measure is:
\begin{equation}
\mu=DU\prod_i dx_i\prod_{i>j}\left(d^{p-1}a_{ij}^{\perp}d^{p-1}a_{ji}^{\perp}\right){\rm exp}\left\{\sum_{i\neq j}\log|\vec\Delta_{ij}|-2\sum_{i\neq j}{\bar{\vec a}}_{ij}.\hat D_{ij}.\vec a_{ji}+O(a^3)\right\}\ ,\label{measure}
\end{equation}
   

\section{The constants $\#({p})$ and $\tilde\#({p})$}
In this section we will provide details about the analytic evaluation of the quantities $\#(p)$ and $\tilde\#(p)$ defined in equations (\ref{numberp}) and (\ref{numberp'}) respectively. Note that one can write:
\begin{equation}
\#(p)=\frac{p-5}{2}A_p+2C_p;~~~\tilde\#(p)=\frac{p-2}{2}A_p; \ , \label{numbers}
\end{equation}
where:
\begin{eqnarray}
A_p&=&\int_{B^p}\int_{B^p}\int_{B^p}d^p\tilde x d^p\tilde yd^p\tilde z \rho(\tilde{\vec x}) \rho(\tilde{\vec y}) \rho(\tilde{\vec z})\frac{1}{(\tilde{\vec x}-\tilde{\vec z} )^2(\tilde{\vec y}-\tilde{\vec z} )^2}\label{ap}\\
C_p&=&\int_{B^p}\int_{B^p}\int_{B^p}d^p\tilde x d^p\tilde yd^p\tilde z \rho(\tilde{\vec x}) \rho(\tilde{\vec y}) \rho(\tilde{\vec z})\frac{\cos^2\theta_{x,yz}}{(\tilde{\vec x}-\tilde{\vec z} )^2(\tilde{\vec y}-\tilde{\vec z} )^2}\label{cp}.
\end{eqnarray}
Let us consider first the quantity $A_p$. Note that the integrals along $\tilde x$ and $\tilde y$ in equation (\ref{ap}) factorize and one can write:
\begin{equation}
A_p=\int_{B^p}d^p\tilde z \rho(\tilde{\vec z})Q_p(|\tilde{\vec z}|)^2=\frac{2\pi^{p/2}}{\Gamma(p/2)}\int\limits_0^1 d\tilde z {\tilde z}^{p-1}\rho(\tilde{z})Q_p(\tilde z)^2\ ,
\end{equation}
where:
\begin{equation}
Q_p(\tilde z)\equiv\int_{B^p}d^p\tilde x \rho(\tilde{x})\frac{1}{(\tilde{\vec x}-\tilde{\vec z} )^2}
\end{equation}
and we have used that the distribution is $SO(p)$ symmetric (namely $\rho(\tilde{\vec x})=\rho(\tilde{x})$). Note that for a uniform distribution we have $\rho(\tilde{x})={p\Gamma(p/2)}/{2\pi^{p/2}}$ and we have:
\begin{equation}
A_p=p\int\limits_0^1 d\tilde z {\tilde z}^{p-1}Q_p(\tilde z)^2\ . 
\end{equation}
One can show that for $p=3$ and a uniform distribution $Q_3(\tilde z)$ is given by:
\begin{equation}
Q_3(\tilde z)=\frac{3}{2}\left[1+\frac{1-{\tilde z}^2}{z}{\rm Arctanh}(\tilde z)\right]
\end{equation}
and
\begin{equation}
A_3=3\int\limits_0^1 d\tilde z {\tilde z}^{2}Q_3(\tilde z)^2=\frac{3}{20}\left(15+2\pi^2\right)\ .\label{a3}
\end{equation}
Let us now focus on the quantity $C_p$. After using equation (\ref{cos}) and the $SO(p)$ symmetry of the eigenvalue distribution, one can show that:
\begin{eqnarray}
C_p&=&\frac{3}{4}A_p-\int_{B^p}\int_{B^p}\int_{B^p}d^p\tilde x d^p\tilde yd^p\tilde z \rho(\tilde{\vec x}) \rho(\tilde{\vec y}) \rho(\tilde{\vec z})\frac{(\tilde{\vec x}-\tilde{\vec z}).(\tilde{\vec y}-\tilde{\vec z} )}{(\tilde{\vec x}-\tilde{\vec z} )^2(\tilde{\vec y}-\tilde{\vec z} )^4}\\
&=&\frac{3}{4}A_p+\frac{2\pi^{p/2}}{\Gamma(p/2)}\int\limits_0^1 d\tilde z {\tilde z}^{p-1}\rho(\tilde{z})Q_p'(\tilde z)\Phi_p'(\tilde z)\ ,
\end{eqnarray}
where
\begin{equation}
\Phi_p(\tilde z)\equiv \frac{1}{2}\int_{B^p}d^p\tilde x\rho(\tilde{x}) \log|(\tilde{\vec x}-\tilde{\vec z})|\ .
\end{equation}
For an uniform distribution $\rho(\tilde z)$ we have:
\begin{equation}
C_p=\frac{3}{4}A_p+p\int\limits_0^1 d\tilde z {\tilde z}^{p-1}Q_p'(\tilde z)\Phi_p'(\tilde z)\ .
\end{equation}
One can show that for $p=3$ and a uniform distribution $\Phi_3(\tilde z)$ is given by:
\begin{equation}
\Phi_3(\tilde z)=\frac{1}{96\tilde z}\left[-34\tilde z+6\tilde z^3+3(\tilde z-1)^3(3+\tilde z)\log|1-\tilde z|-3(\tilde z-3)(1+\tilde z)^3\log|1+\tilde z|\right]
\end{equation}
and one can calculate $C_3$:
\begin{equation}
C_3=\frac{3}{20}\left(15+\frac{1}{2}\pi^2\right)\label{c3}\ .
\end{equation}
Now one can substitute the results from equations (\ref{a3}) and (\ref{c3}) into equation (\ref{numbers}) to obtain:
\begin{equation}
\#(3)=\frac{3}{20}(15-\pi^2);~~~\tilde\#(3)=\frac{3}{40}(15+2\pi^2); \ , \label{numberscal}
\end{equation}

\end{document}